\documentstyle[preprint,aps]{revtex}
\begin{document}
\draft
\preprint{
\parbox{4cm}{
\baselineskip=12pt
TMUP-HEL-9711\\ 
November, 1997\\
Revised 2/2/98
\hspace*{1cm}
}}
\title{ A Simple Model of Gauge Mediated Supersymmetry Breaking \\
        with Composite Messenger Fields}
\author{ Nobuchika Okada 
 \thanks{e-mail: n-okada@phys.metro-u.ac.jp}
\thanks{JSPS Research Fellow}}
\address{Department of Physics, Tokyo Metropolitan University,\\
         Hachioji-shi, Tokyo 192-03, Japan}
\maketitle
\vskip 2.5cm
\begin{center}
{\large Abstract}
\vskip 0.7cm
\begin{minipage}[t]{14cm}
\baselineskip=19pt
\hskip4mm
We present a simple model of gauge mediated supersymmetry 
breaking with composite messenger fields. 
Our model is based on the gauge group $SP(8) \times SU(2)$. 
By the strong $SP(8)$ dynamics, supersymmetry is dynamically broken 
 and the composite fields with charges under the standard model gauge group 
 appear at low energy. 
The $U(1)_R$ symmetry breaking mass terms for the composite fields are 
 generated by the strong $SU(2)$ dynamics. 
Then, the composite fields play a role of the messenger fields. 
On the other hand, the theoretical bounds on the parameters in our model 
 are discussed. 
Especially, the lower bound on the dynamical scale 
 of the $SP(8) \times SU(2)$ gauge interaction is roughly $10^{15}$ GeV.
\end{minipage}
\end{center}
\newpage

The models of gauge mediated supersymmetry breaking (GMSB) 
 have attractive feature 
 in the minimal supersymmetric standard model (MSSM). 
Since supersymmetry breaking is mediated to the MSSM sector 
 by the standard model gauge interaction 
 through the messenger fields which are charged 
 under the MSSM gauge group, 
 the superpartners with the same charges in the MSSM get 
 the same soft supersymmetry breaking masses.  
As a result, the problem of the flavor changing neutral current 
 in the MSSM are resolved naturally. 

The pioneering works have been done by Dine, Nelson
 and co-workers \cite{dine}. 
They have constructed explicit models which realized the mediation of 
 supersymmetry breaking to the MSSM sector. 
Furthermore, it has been shown that the models was phenomenologically viable. 
 
However, the original models were very complicated. 
This fact originates from the complexity 
 of the dynamical supersymmetry breaking mechanism. 
In addition, introduction of three separated sectors, 
 the supersymmetry breaking sector, the messenger sector 
 and the MSSM sector, make the models more complicated. 

Several attempts to obtain more simple GMSB models have been considered 
 by many authors. 
A simple mechanism of the dynamical supersymmetry breaking 
 has been proposed by Izawa and Yanagida, 
 and Intriligator and Thomas \cite{Yanagida},  
 and this mechanism was applied to the supersymmetry breaking sector 
 in the GMSB models \cite{hotta}.  
Moreover, new types of the GMSB models 
 in which the messenger sector is unified 
 into the supersymmetry breaking sector 
 have been constructed \cite{composite}. 

In this letter, we present a simple GMSB model 
 based on the gauge group $SP(8) \times SU(2)$.    
Supersymmetry is dynamically broken by the strong $SP(8)$ gauge dynamics. 
Since the standard model gauge group 
 $SU(5)_{SM}\supset SU(3)_c \times SU(2)_L \times U(1)_Y$ is embedded 
 in the global symmetry $SU(10)$ which the $SP(8)$ gauge dynamics has, 
 the messenger sector is unified into the supersymmetry breaking sector 
 and the messenger fields appear as composite fields at low energy. 
The strong $SU(2)$ gauge dynamics generates 
 the $U(1)_R$ symmetry breaking mass terms 
 for the messenger fields. 

Before discussing our model, 
 let us review the messenger sector. 
The typical superpotential is simply described by 
\begin{eqnarray}
W_{mess} = \sum_i \lambda_i Z_i \bar{\Phi} \Phi \; \; , 
\label{mess}
\end{eqnarray}
where $\bar{\Phi}$ and $\Phi$ have the vector-like charge 
 under the MSSM gauge group, $Z_i$ is a singlet field 
 under the gauge group,  
 and $\lambda_i$ is a dimensionless coupling constant. 
If nonzero vacuum expectation values 
 of the $F$-component of at least one $Z_i$ 
 and the scalar component of at least one $Z_j$ are realized, 
 the fields $\bar{\Phi}$ and $\Phi$ 
 can play a role of the messenger fields. 
Note that $i=j$ is not needed in general.  

Our model is based on the gauge group $SP(8) \times SU(2)$ as mentioned above. 
To make our discussion clear, let us consider only the $SP(8)$ dynamics 
 at first. 
The particle contents are as follows. 
\begin{center}
\begin{tabular}{cccc}
 \hspace{1cm}    & $~SP(8)~$ & $~SU(5)_{SM}~$          & $~U(1)_R~$   \\
\hline 
$ \bar{P} $   & \bf{8}      & $\bar{\bf{5}}$ &  0  \\
$ P  $             & \bf{8}      & \bf{5}              &  0   \\
\hline
$ Z  $             & \bf{1}      & \bf{1}              &  2  \\
$ Z^{\prime}$             & \bf{1}      & \bf{1}         &  2  \\
$ \bar{\phi}$ & \bf{1}      & $\bar{\bf{10}}$    &  2  \\
$ \phi $           & \bf{1}      & \bf{10}        &  2  \\
$  \bar{A} $  & \bf{1}      & \bf{24}        &  2  \\
$ N  $             & \bf{1}      &  \bf{1}        &  0
\end{tabular}
\end{center}
Note that the standard model gauge group 
 $SU(5)_{SM} \supset SU(3)_c \times SU(2)_L \times U(1)_Y$ is embedded 
 in the global symmetry $SU(10)$ which the $SP(8)$ dynamics has.  
In this paper we always use the notation of the ordinary $SU(5)$ 
 Grand Unified Theory (GUT), for simplicity. 
It is trivial to decompose it into the standard model notation.  

A renormalizable tree level superpotential 
 which is consistent with all the symmetry is given by 
\footnote{
We assume that all of the parameters in our model 
 are real and positive, for simplicity.} 
\begin{eqnarray}
   W_{tree}= \lambda_Z Z \; \left[\bar{P} P\right]_s 
  + \lambda_{Z^\prime}  Z^\prime ( \left[\bar{P}P\right]_s - \lambda_N N^2 ) 
 \nonumber \\
   + \lambda_{\bar{\phi}} \bar{\phi} \; \left[PP \right] 
   + \lambda_{\phi} \phi \; \left[\bar{P}  \bar{P}\right] 
   + \lambda_{\bar{A}}\; 
   \mbox{tr} (\bar{A}\;  \left[\bar{P} P\right]_{adj}) \; \; , 
 \label{tree1}
\end{eqnarray}
where square brackets denote the contraction of the $SP(8)$ indices,   
 and $\left[\; \right]_s$ and $\left[ \; \right]_{adj}$ denote 
 to extract a part of singlet and adjoint representation 
 of $SU(5)_{SM}$ from $\left[ \bar{P} P \right]$, respectively. 
Here, we assume that the tree level superpotential has no 
 dimensionful parameter. 
As can be seen in the following, in our model, 
 all of the dimensionful parameters are dynamically generated  
 and originate from strong gauge dynamics. 

We can obtain the low energy description of this theory 
 by the method of Seiberg and co-workers \cite{seiberg}. 
The moduli space is dynamically deformed  
 to satisfy the condition Pf$V=\Lambda^{10}$,  
 where $V$ is $10 \times 10$ antisymmetric tensor given by 
\begin{eqnarray}
   V =  \left[ 
   \begin{array}{cc}
    \left[\bar{P} \bar{P}\right]    &  \left[\bar{P} P \right] \\
    \left[P \bar{P}\right]          & \left[ P P \right]  \\
   \end{array}   \right]  
  \sim  \Lambda \left[
   \begin{array}{cc}
     \bar{\Phi}    &  S + A \\
     -S - A        &  \Phi  \\
   \end{array}     \right]  \; \; . 
\end{eqnarray}
Here, $ \Lambda $ is the dynamical scale of the $SP(8)$ gauge interaction.  
The fields $S$, $A$, $\bar{\Phi}$ and $\Phi$ 
 are the effective fields as follows. 
\begin{center}
 \begin{tabular}{cclc}
\hspace{1cm} &  \hspace{1cm}   &  \hspace{1cm} & $ ~SU(5)_{SM}~ $ \\
$ S $ & $\sim $ & $\left[\bar{P}P\right]_s/\Lambda $  &  \bf{1}      \\
$ A $ & $\sim $ &$ \left[\bar{P}P\right]_{adj}/\Lambda$& \bf{24}      \\
$ \bar{\Phi} $ & $\sim $ & $\left[\bar{P} \bar{P}\right]/\Lambda $ 
& $\bf{\bar{10}}$ \\ 
$ \Phi $ & $\sim $ & $ \left[PP\right]/\Lambda $    & \bf{10} 
 \end{tabular}
\end{center}
Since the condition Pf$V=\Lambda^{10}$ contradicts 
 the supersymmetric vacuum conditions 
 required by the tree level superpotential of eq.(\ref{tree1}), 
 supersymmetry is dynamically broken \cite{Yanagida}. 

To obtain the effective superpotential at low energy, 
 we should eliminate one of the effective fields 
 by considering the condition Pf$V=\Lambda^{10}$.   
Using the effective fields, the condition is described by 
\begin{eqnarray}
 S^5 - S^3 \left( \bar{\Phi} \Phi + \frac{1}{2} \mbox{tr}A^2\right) 
 + \frac{1}{3}S^2 \mbox{tr}A^3 
 + S \left\{ \left(\bar{\Phi} \Phi\right)^2 - \frac{1}{4} 
 \mbox{tr} A^4 \right\}
 -\Phi^2 A {\bar{\Phi}}^2 + \frac{1}{5} \mbox{tr} A^5 
 = \Lambda^5 \; \; .
\end{eqnarray}
Considering small fluctuation of $S$ around $\langle S \rangle = \Lambda$, 
 we can obtain 
\begin{eqnarray}
     S \sim \Lambda + \frac{1}{5 \Lambda} 
    \left( \bar{\Phi} \Phi + \frac{1}{2} \mbox{tr} A^2 \right) 
\end{eqnarray}
Eliminating $S$ from eq.(\ref{tree1}), 
 the effective superpotential is given by 
\begin{eqnarray}
   W_{eff} &\sim & \lambda_Z Z 
   \left\{ \Lambda^2+ \frac{1}{5}
   \left( \bar{\Phi} \Phi + \frac{1}{2} \mbox{tr} A^2 \right) \right\}
    \nonumber \\
   &+&  \lambda_{Z^\prime} Z^\prime \left\{ \Lambda^2 + \frac{1}{5}
   \left( \bar{\Phi} \Phi + \frac{1}{2} \mbox{tr} A^2 \right) 
    - \lambda_N N^2  \right\}  \nonumber \\
   & + & \lambda_{\bar{\phi}}\Lambda \; \bar{\phi} \Phi
   + \lambda_{\phi} \Lambda \; \bar{\Phi} \phi
   + \lambda_{\bar{A}} \Lambda \; \mbox{tr} (\bar{A}A) \; \; . \label{eff1}
\end{eqnarray}
This effective superpotential is 
 one of the type of O'Raifeartaigh model \cite{O'R}. 
For small value of $\lambda_Z$ compared with 
 $\lambda_{\bar{\phi}}$, $\lambda_{\phi}$ and $\lambda_{\bar{A}}$, 
 supersymmetry is broken by $\langle F_Z \rangle = - \lambda_Z \Lambda^2$, 
 where $ F_Z $ is the $F$-component of $Z$. 

However, note that the scalar potential derived from eq.(\ref{eff1}) has
 the `pseudo-flat' direction, namely, 
 the potential remains minimum along arbitrary value of $\langle Z \rangle$
\footnote{ We use the same notation for the superfield itself 
 and the scalar component of the superfield.  
}. 
This `pseudo-flat' direction is lifted up by quantum corrections 
 for the effective potential of $Z$.   
There are two possibilities where the effective potential has minimum. 
One is $\langle Z \rangle \sim \Lambda$ which may be expected 
 by the effect of the strong $SP(8)$ interaction 
 \cite{Yanagida}\cite{hotta}. 
The other is $\langle Z \rangle =0 $ which is expected 
 only if the Yukawa coupling in eq.(\ref{eff1}) is considered \cite{hqu}. 
Unfortunately, there is currently no technique to definitely decide 
 which vacuum is chosen. 
In this letter, we assume that true vacuum lies at $\langle Z \rangle =0$. 

Then, the vacuum is realized at 
 $\langle F_Z \rangle \neq 0$, $\langle$other $F$-components$\rangle =0$, 
 $\langle N \rangle = \Lambda / \sqrt{\lambda_N}$, 
 and $\langle$other scalar components$\rangle =0$. 
Note that there is no $U(1)_R$ symmetry breaking mass term 
 for $\bar{\Phi}$, $\Phi$ and $A$ in the effective superpotential, 
 because of $\langle Z \rangle = \langle Z^\prime \rangle =0$.  
Therefore, the fields $\bar{\Phi}$, $\Phi$ and $A$ cannot play a role of 
 the messenger fields. 
For example, the gauginos in the MSSM cannot get 
 their soft supersymmetry breaking masses, 
 since the masses are protected by the $U(1)_R$ symmetry.  

In order to generate the $U(1)_R$ symmetry breaking mass terms 
 for the fields $\bar{\Phi}$, $\Phi$ and $A$, 
 we introduce new strong $SU(2)$ gauge interaction 
 with two doublet fields $\bar{Q}$ and $Q$ 
 which are singlets of $SU(5)_{SM}$. 
In addition to the effective superpotential of eq.(\ref{eff1}), 
 let us consider new tree level superpotential 
\begin{eqnarray}
 W^{\prime}_{tree} = \lambda_M N \; \left[ \bar{Q} Q  \right] \; \; ,
\label{tree2}
\end{eqnarray} 
where $\left[ \; \right]$ denotes the contraction of the $SU(2)$ indices 
 by the $\epsilon$-tensor. 
Although this superpotential is the simplest one to attain our aim, 
 the $U(1)_R$ symmetry is explicitly broken by the $SU(2)$ gauge anomaly. 
This may suggest that a modification of our model is needed. 
However, there is no R-axion problem because of this explicit breaking.  
The vacuum is realized with the same vacuum expectation values of 
 the scaler fields discussed above 
 and $\langle \bar{Q} \rangle = \langle Q \rangle =0$. 

However, we should take into account the non-perturbative effect of 
 the strong $SU(2)$ gauge interaction at low energy. 
When the effect is considered, 
 the effective superpotential is given by \cite{ads} 
\begin{eqnarray}
 W^{\prime}_{eff}= \lambda_M \Lambda^{\prime} N M 
  + \frac{\Lambda^{\prime 4}}{M} \; \; ,  
\label{eff2} 
\end{eqnarray}
where $\Lambda^{\prime}$ is the dynamical scale 
 of the $SU(2)$ gauge interaction, 
 and $M \sim \left[\bar{Q}Q \right] / \Lambda^{\prime}$ 
 is the effective fields. 
Now we obtain the effective superpotential 
 $\tilde{W}_{eff}= W_{eff}+W^{\prime}_{eff}$ 
 as the total effective superpotential in our model.  

Let us investigate where the vacuum is realized. 
 The vacuum is changed and $\langle Z^{\prime} \rangle \neq 0$ occur  
 by the strong $SU(2)$ dynamics. 
Indeed, from two conditions  $\partial \tilde{W}_{eff}/ \partial M = 0$ 
 and $\partial \tilde{W}_{eff}/ \partial N = 0$, we obtain 
\begin{eqnarray}
 \langle M \rangle &=& \sqrt{\frac{1}{\lambda_M \langle N \rangle}}
 \Lambda^{\prime 3/2} 
 = \frac{\lambda_N^{1/4}}{\lambda_M^{1/2}}  
   \frac{\Lambda^{\prime 3/2}}{\Lambda^{1/2}} \; \; ,  \nonumber \\
 \langle Z^{\prime} \rangle &=& 
 \frac{\lambda_M}{2 \lambda_{Z^\prime} \lambda_N} 
 \frac{\langle M \rangle}{\langle N \rangle} \Lambda^{\prime} 
 = \frac{\lambda_M^{1/2}}{2 \lambda_{Z^\prime}\lambda_N^{1/4}}
   \frac{\Lambda^{\prime 5/2}}{\Lambda^{3/2}}  \; \; . 
 \label{vac}
\end{eqnarray}
Then, the $U(1)_R$ symmetry breaking mass terms 
 for the fields $\bar{\Phi}$, $\Phi$ and $A$ are generated. 
The effective superpotential corresponding to eq.(\ref{mess})
 is described by 
\begin{eqnarray}
W_{mess} = \frac{1}{5} (\lambda_Z Z + \lambda_{Z^\prime} Z^\prime)
   \left( \bar{\Phi} \Phi + \frac{1}{2} \mbox{tr} A^2 \right) \; \; . 
\label{mess2}
\end{eqnarray}
Because of $\langle F_Z \rangle \neq 0 $ 
  and $ \langle Z^\prime \rangle \neq 0 $, 
  the composite fields $\bar{\Phi}$, $\Phi$ and $A$ can 
  play a role of the messenger fields. 

The mass spectra of all the superpartners in the MSSM are 
 calculated by this superpotential \cite{martin} 
 with $\langle F_Z \rangle$ and $ \langle Z^\prime \rangle $. 
The gauginos get their soft supersymmetry breaking masses 
 through the one-loop radiative correction by the messenger fields 
 $\bar{\Phi}$, $\Phi$ and $A$. 
For simplicity, let us assume $\lambda_Z \langle F_Z \rangle \ll 
 ( \lambda_{Z^\prime} \langle Z^\prime \rangle)^2 $ and 
 $ \lambda_{\bar{\phi}}\Lambda \sim  \lambda_{\phi}\Lambda \sim 
 \lambda_{\bar{A}}\Lambda \ll \lambda_{Z^\prime} \langle Z^\prime \rangle $. 
Then, the masses of the gauginos are given by 
\begin{eqnarray}
  m_{\lambda_a} \sim \frac{\alpha_a}{4 \pi} 
   \frac{\lambda_Z \langle F_Z \rangle}
      {\lambda_{Z^\prime} \langle Z^\prime \rangle}
   \sum_i n_a(i)  \; \; ,  
\label{gaugino} 
\end{eqnarray}
where $a=1,2$ and $3$ correspond to the MSSM gauge interaction, 
 $SU(3)_c$, $SU(2)_L$ and $U(1)_Y$, respectively, 
 and $n_a(i)$ is the Dynkin index for the messenger fields running the loop, 
 which is defined as $n_a(i)=1$ for $i=\bf{N+\bar{N}}$ of $SU(N)$ 
 and $n_1=6/5 Y^2$ for the messenger fields with the hypercharge $Y$ 
 by using the $SU(5)$ GUT normalization. 
Since the messenger fields have the charge 
 $\bf{10+\bar{10}}$ and $\bf{24}$ of $SU(5)_{SM}$, 
 $\sum_i n_3 = \sum_i n_2 = \sum_i n_1 = 8 $. 
The scalar partners in the MSSM get their masses 
 through the two-loop radiative correction. 
They are given by 
\begin{eqnarray}
  \tilde{m}^2  \sim 2 \left( \frac{\alpha_a}{4 \pi} \right)^2 
   \left( \frac{\lambda_Z \langle F_Z \rangle}
      {\lambda_{Z^\prime} \langle Z^\prime \rangle}  \right)^2
   (\sum_a C_a)\; ( \sum_i n_a(i)) \; \; ,   
\label{scalar}
\end{eqnarray}
where $C_a$ is the quadratic Casimir invariant for the scalar partners  
 which is defined as $C_3=4/3$, $C_2=3/2$ and $C_1=3/5 Y^2$. 
If the values of parameters $\lambda_Z$, $\lambda_M$, $\lambda_N$, 
 $\Lambda$ and $\Lambda^\prime$ are fixed, 
 the masses of all the superpartners are fixed 
 by eqs.(\ref{gaugino}) and (\ref{scalar}).  
 
However, all of the values of these parameters 
 are not allowed. 
For simplicity, we take $\lambda_M \sim \lambda_N \sim \cal{O}$(1) and 
 $\Lambda = \Lambda^\prime$. 
Then, the dynamical scale $\Lambda$ has a theoretical lower bound. 
Since there are many charged particles in addition to the ordinary 
 quarks and leptons in our model, 
 the QCD gauge coupling blows up below the Planck scale, 
 unless the dynamical scale of the $SP(8) \times SU(2)$ gauge interaction 
 is high enough. 
We define mass scale of the fields $\bar{\phi}$, $\phi$ and $\bar{A}$ 
 as $m^\prime = \lambda_{\bar{\phi}}\Lambda \sim  \lambda_{\phi}\Lambda 
 \sim \lambda_{\bar{A}}\Lambda$, 
 and the messenger scale as $m = \lambda_{Z^\prime} \langle Z^\prime \rangle 
 \sim 1/2 \; \Lambda$. 
Let us consider one-loop renormalization group equation (RGE) 
 of the QCD coupling \cite{abh}. 
At the scale $M_{SUSY} \leq \mu \leq m^\prime$, 
 the solution to the RGE is given by
\begin{eqnarray} 
 \frac{1}{\alpha_3(M_{SUSY})} -  \frac{1}{\alpha_3(\mu)}= 
 -\frac{3}{2 \pi} \ln (\mu/M_{SUSY}) \; \; , 
\label{RGE1}
\end{eqnarray}
where $M_{SUSY} \sim 1$TeV is a typical value of masses of 
 the superpartners in the MSSM. 
At the scale $m^\prime \leq \mu \leq m$ 
 (remember our assumption $m^\prime \ll m$), 
 the fields $\bar{\phi}$, $\phi$ and $\bar{A}$ contribute to the RGE, 
 and the solution is given by
\begin{eqnarray}
 \frac{1}{\alpha_3(m^\prime)} -  \frac{1}{\alpha_3(\mu)}=
 \frac{5}{2 \pi} \ln (\mu/m^\prime) \; \; .
\label{RGE2}
\end{eqnarray}
At the scale $m \leq \mu$ where all of the colored fields 
 contribute to the RGE, we obtain 
\begin{eqnarray}
 \frac{1}{\alpha_3(m)} -  \frac{1}{\alpha_3(\mu)}=
 \frac{13}{2 \pi} \ln (\mu/m) \; \; . 
\label{RGE3}
\end{eqnarray}
Note that this solution is not changed at $ \Lambda \leq \mu $ 
 where the dynamical degrees of freedom of the messenger fields 
 are replaced by that of the elementary fields $\bar{P}$ and $P$. 
Let us define the theoretical lower bound on $m=1/2 \; \Lambda$ 
 as $1/ \alpha(M_{Pl}) =0$, 
 where $M_{Pl}=10^{19}$GeV is the Planck scale.  
From eqs.(\ref{RGE1}), (\ref{RGE2}) and (\ref{RGE3}), 
 the bound is given by 
\begin{eqnarray}
m =  \delta^{-1/2} \; M^{3/16}_{SUSY}\; 
  M_{Pl}^{13/16}  \Bigg/  
  \exp \left( \frac{\pi}{8 \alpha_3(M_{SUSY})} \right) 
 \sim \delta^{-1/2} \; 10^{14} \; \; \mbox{GeV} \; \; , 
 \label{eq16} 
\end{eqnarray}
where $\delta$ is defined as $\delta= m^\prime / m$, and  
 we take $1/\alpha_3(M_{SUSY})\sim 12$.   
If we take $\delta \sim 10^{-2}$, the lower bound on the 
 dynamical scale of the $SP(8) \times SU(2)$ gauge interaction 
 is given by $\Lambda \sim 10^{15}$ GeV. 

Next, let us investigate the upper bound on $\lambda_Z$ 
 by implying the naturalness criterion \cite{thooft}. 
According to the criterion, the masses of the scalar partners 
 in the MSSM should be less than 1 TeV. 
From eq.(\ref{scalar}), we obtain 
\begin{eqnarray}
\frac{2 \alpha_3}{\sqrt{3}\pi}\; \frac{\lambda_Z \langle F_Z \rangle}
         {\lambda_{Z^\prime}\langle  Z^\prime \rangle} 
 \sim \frac{4 \alpha_3}{\sqrt{3}\pi}\; 
 \lambda_Z^2 \; \Lambda  \leq 1 \; \; 
  \mbox{TeV} \; \; , 
 \label{eq17}
\end{eqnarray}
where $C_3=4/3$ and $\sum_i n_3 = 8$ are used. 
Considering the lower bound on $\Lambda \geq 10^{15}$ GeV, 
 the upper bound on $\lambda_Z \leq 10^{-6}$
 is obtained, where we take $\alpha_3 \sim 0.1$. 
Note that this upper bound is consistent 
 with our assumption $\lambda_Z \langle F_Z \rangle \ll m^2$ 
 used to obtain eqs.(\ref{gaugino}) and (\ref{scalar}).  

Here, we give a comment on the value of $\lambda_Z$. 
Although the upper bound on $\lambda_Z \leq 10^{-6}$ seems to be 
 unnaturally small, this result is due to our assumption 
 $\Lambda = \Lambda^\prime$, and can be avoided 
 in the case $\Lambda \ll \Lambda^\prime$. 
Eqs.(\ref{vac}) and (\ref{eq17}) suggest that the upper bound of $\lambda_Z$ 
 becomes larger as $\Lambda^\prime$ becomes larger than $\Lambda$.  
For example, if we take $\Lambda=4 \times 10^9$ and 
 $\Lambda^\prime=6 \times 10^{11}$ which satisfy eq.(\ref{eq16}), 
 $\lambda_Z \leq {\cal O}(1)$ can be obtained 
 from eqs.(\ref{vac}), (\ref{eq16}) and (\ref{eq17}). 

In summary, we present a simple model of the gauge mediated 
 supersymmetry breaking. 
Our model is based on the gauge group $SP(8) \times SU(2)$. 
Supersymmetry is dynamically broken by the strong SP(8) dynamics, 
 and the composite fields which would be the messenger fields 
 also appear by this dynamics. 
At this stage, there is no $U(1)_R$ symmetry breaking mass term  
 for the composite fields. 
The mass terms are generated by the strong $SU(2)$ dynamics. 
Then, the composite fields can play a role of the messenger fields. 
On the other hand, the theoretical bounds on the parameters in our model
 is discussed. 
The dynamical scale of the $SP(8) \times SU(2)$ gauge interaction 
 should be more than $10^{15}$ GeV  
 to prevent the QCD coupling from blowing up below the Planck scale. 
The naturalness criterion requires $\lambda_Z \leq 10^{-6}$ 
 together with the lower bound on the dynamical scale. 

Finally, we would like to comment on a possibility of extension of our model. 
The gauge group $SP(8)$ is minimal one 
 to be able to include fields 
 with the vector-like $\bf{5+\bar{5}}$ representation 
 under the MSSM gauge group into the $SP(8)$ dynamics. 
It is possible to introduce the vector-like fields,  
 only if the number of flavors is more than five. 
Therefore, we can extend the gauge group $SP(8)$ to $SP(2N)$ ($N \geq 5$) 
 with $N+1$ flavors in general. 
On the other hand, the gauge group $SU(2)$ is also minimal one. 
It is possible to generate the $U(1)_R$ symmetry breaking mass terms 
 for the messenger fields by the same mechanism discussed above, 
 only if $N_f < N_C$, 
 where $N_f$ and $N_C$ are number of flavors and colors of $SU(N_C)$,
 respectively. 
Therefore, we can extend the gauge group $SU(2)$ to $SU(N)$ ($N \geq 3$)
 with $N_f < N$ flavors in general. 

The author would like to thank Noriaki Kitazawa for useful comments. 
This work was supported in part by the Grant 
in Aid for Scientific Research from the Ministry of Education, 
Science and Culture and the Research Fellowship of the Japan Society 
for the Promotion of Science for Young Scientists. 
%


\begin{references}
\bibitem{dine}
 M. Dine and A.E. Nelson,
 Phys. Rev. D 48 (1993) 1277;
 M. Dine, A.E. Nelson and Y. Shirman,
 Phys. Rev. D 51 (1995) 1362;
 M. Dine, A.E. Nelson, Y. Nir and Y. Shirman,
 Phys. Rev. D 53 (1996) 2658. 
%
\bibitem{Yanagida}
 Izawa K.-I.  and T. Yanagida,
 Prog. Theor. Phys. 95 (1996) 949; 
 K. Intriligator and S. Thomas,
 Nucl. Phys. B 473 (1996) 121. 
%
\bibitem{hotta}
 T. Hotta, Izawa K.-I. and T. Yanagida,
 Phys. Rev. D 55 (1997) 415; 
 Izawa K.-I., preprint hep-ph/9704382; 
 Izawa K.-I., Y. Nomura, K. Tobe and T. Yanagida,
 Phys. Rev. D 56 (1997) 2886. 
%
\bibitem{composite}
 E. Poppitz and S.P. Trivedi,
 Phys. Rev. D 55 (1997) 5508;
 H. Murayama,
 Phys. Rev. Lett 79 (1997) 18;
 S. Dimopoulos, G. Dvali and R. Rattazzi, 
 preprint hep-ph/9707537; 
 M. Luty and J. Terning,
 preprint hep-ph/9709306; 
  Y. Shirman,
 preprint hep-ph/9709383. 
%
\bibitem{seiberg}
 N. Seiberg,
 Phys. Lett. B 318 (1993) 469;
 N. Seiberg,
 Phys. Rev. D 49 (1994) 6857;
 N. Intriligator, R.G. Leigh and N. Seiberg,
 Phys. Rev. D 50 (1994) 1092.
%
\bibitem{O'R}
 L. O'Raifeartaigh,
 Nucl. Phys. B 96 (1975) 331.
%
\bibitem{hqu}
 M. Hqu, 
 Phys. Rev. D 14 (1976) 3548. 
%
\bibitem{ads} 
 I. Affleck, M. Dine and N. Seiberg, 
 Nucl. Phys. B 256 (1985) 557. 
%
\bibitem{martin}
 S.P. Martin, 
 Phys. Rev. D 55 (1997) 3177.
%
\bibitem{abh}
 U. Amaldi, W.de Boer and H. F\"urstenau, 
 Phys. Lett. B 260 (1991) 447.
%
\bibitem{thooft}
 G. 't Hooft, 1979 Cargese Lectures, 
 published in Recent Developments In Gauge Theories, 
 Proceedings, NATO Advanced Study Institute New York, 
 USA: Plenum (1980).
\end{references}
\end{document}